\documentclass[prb,aps,twocolumn,amsmath,amssymb,floatfix,superscriptaddress]{revtex4}
\usepackage{amsmath}
\usepackage{amsfonts}
\usepackage{amssymb}
\usepackage[dvips]{graphics}
\usepackage{color}
\definecolor{dred}{rgb}{0.75,0,0}
\usepackage{soul}
\usepackage[colorlinks=true, citecolor=blue, urlcolor=blue]{hyperref}

\date{\today}

\begin{document}
	
\title{Bias induced circular current in a loop nanojunction with AAH modulation: Role of hopping dimerization} 

\author{Moumita Mondal}

\email{moumitamondal$_$r@isical.ac.in}

\affiliation{Physics and Applied Mathematics Unit, Indian Statistical Institute, 203 Barrackpore Trunk Road, Kolkata-700 108, India}

\author{Santanu K. Maiti}

\email{santanu.maiti@isical.ac.in}

\affiliation{Physics and Applied Mathematics Unit, Indian Statistical Institute, 203 Barrackpore Trunk Road, Kolkata-700 108, India}

\begin{abstract}
	
In this work, we put forward, for the first time, the interplay between correlated disorder and hopping dimerization on bias driven
circular current in a loop conductor that is clamped between two electrodes. The correlated disorder is introduced in site energies of
the ring in the form of Aubry-Andr\'e-Harper (AAH) model. Simulating the quantum system within a tight-binding framework all the results 
are worked out based on the standard wave-guide theory. Unlike transport current, commonly referred to drain current, circular current 
in the loop conductor can get enhanced with increasing disorder strength. This enhancement becomes much effective when hopping dimerization 
is included which is taken following the Su-Schrieffer-Heeger (SSH) model. The characteristic features of bias driven circular current 
are studied under different input conditions and we find the results are robust for wide range of physical parameters. Our analysis 
may provide a new insight in analyzing transport behavior in different disordered lattices in presence of additional restrictions in 
hopping integrals.

\end{abstract}

\maketitle

\section{Introduction}

The generation of circular current due to voltage bias in a nanojunction possessing single and/or multiple loops is relatively a new 
phenomenon~\cite{cr1,cr2,cr3,cr4} compared to the well established non-decaying circular current in an isolated loop in presence 
of magnetic flux~\cite{p1,p2,p3,p4,p5,p6,p7,p8,p9,p10,p11,p12}. For the 
latter case, the magnetic flux is responsible to break the symmetry among clockwise and counter-clockwise propagating electronic waves 
which yields a net circular charge current that even persists when the flux is removed. This phenomenon is referred to as flux-driven 
circular current in an isolated loop conductor (not coupled to external baths) and has been extensively studied in the literature over 
the past several years~\cite{p1,p2,p3,p4,p5,p6,p7,p8,p9,p10,p11,p12}.

When we talk about current in a nanojunction, we essentially consider transport current i.e., the current measured in drain electrode. 
But along with transport current, a circular current can be established~\cite{cr1,cr2,cr3,cr4} in a loop conductor that is directly 
clamped to electronic 
baths under suitable conditions. Here the current is generated due to voltage bias and no magnetic flux is required unlike isolated 
quantum loop. The study of this phenomenon, called as bias driven circular current, leads to several important physical pictures as it 
directly involves current distribution in different segments of a conductor. A very little progress has been made so far by a few 
groups including us considering some molecular junctions as well as different tailored geometries~\cite{cr1,cr2,cr3,cr4}. These 
works have not investigated 
the effect of disorder in a deeper level, though for junction current the role of disorder is well known. Thus, the specific role of 
disorder on bias driven circular current is highly required, and it is one of our primary motivations behind the present study. In 
condensed matter systems different kinds of disorder are taken into account. Among them, correlated disorder in the form of 
Aubry-Andr\'e-Harper models~\cite{aa1,aa2,aa3,aa4,aa5,aa6,aa7,aa8,aa9} nowadays draws significant attention due to its unique and diverse 
characteristic features, compared to usual random disordered lattices~\cite{ran1,ran2,ran3,ran4}. The response of circular current 
on disorder is quite different. Unlike junction current, the circular 
current can get enhanced with increasing the disorder strength. This phenomenon becomes more effective when a restriction is imposed in 
hopping integrals. Instead of identical hopping in all bonds, if we take dimerized hopping scenario, following the Su-Schrieffer-Heeger 
(SSH) model~\cite{ss1,ss2,ss3,ss4,ss5,ss6,ss7}, then much higher circular current can be achieved. 
\begin{figure}[ht]
\centering \resizebox*{8.25cm}{4.25cm}{\includegraphics{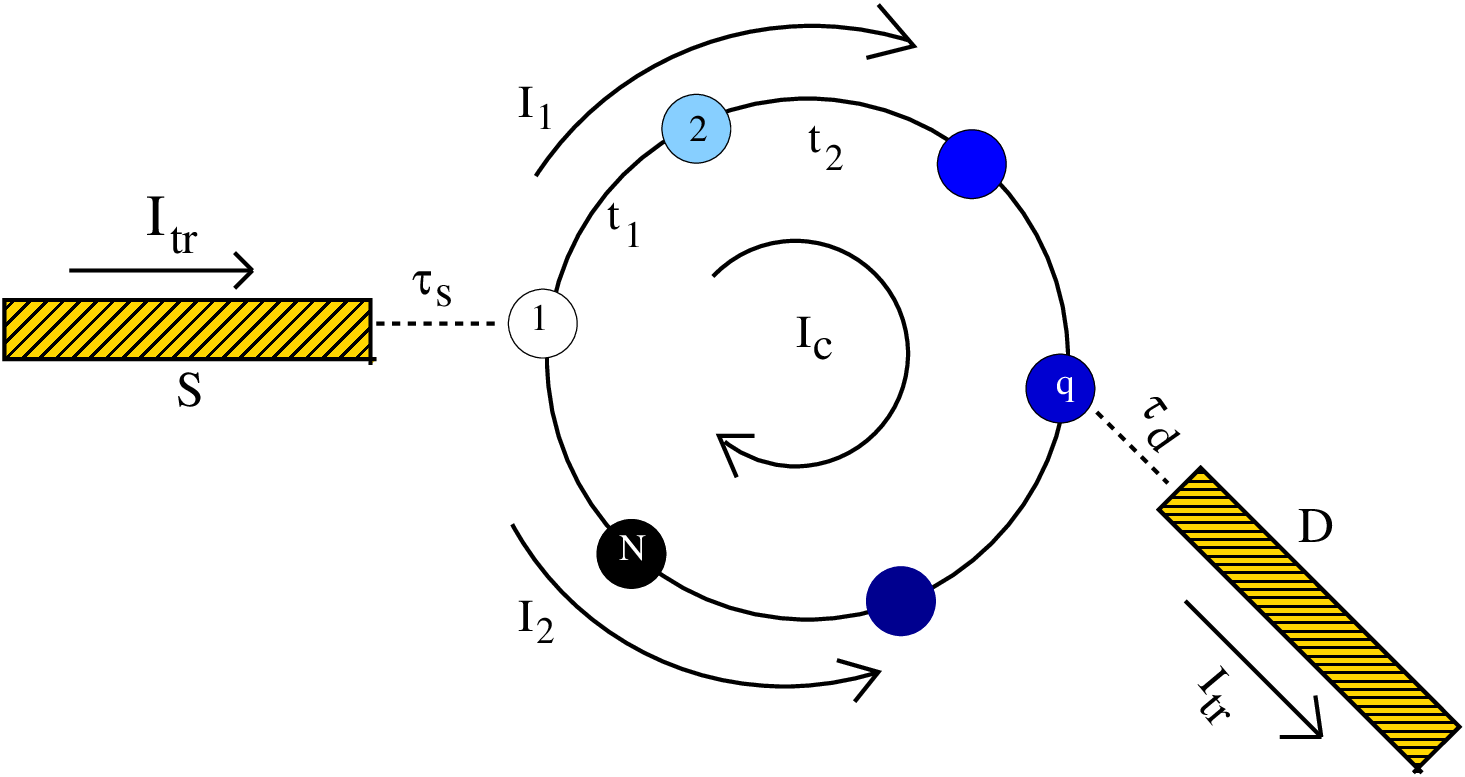}}
\caption{(Color online). Schematic diagram of a ring nanojunction. A circular current $I_c$ is established in the ring when a bias 
is applied among the electrodes, source (S) and drain (D).} 
\label{fig1}
\end{figure}
The combined effect of AAH correlation and hopping dimerization brings several interesting features which we want to investigate in 
the present work. Such an effect has not been discussed so far to the best of our concern and might be helpful in analyzing transport 
phenomena in nanoscale regime.

We take a nanojunction where a single ring is coupled to two electronic baths (see Fig.~\ref{fig1}). A tight-binding (TB) framework is 
given to describe the nanojunction, and, all the results are worked out based on the standard wave-guide theory. Different aspects are
taken into account which include the ring-electrode junction configuration, Fermi energy, system size, AAH phase and related issues.
The key findings are 
(i) observation of higher circular current with disorder, (ii) more enhancement can occur with the inclusion of hopping dimerization, 
and (iii) all the results are valid for broad range of physical parameters.

The rest part of our work is organized as follows. Section II includes the description of model ring nanojunction along with the 
required theoretical framework to compute the results. All the essential results are presented and critically analyzed in Sec. III. 
Finally, Sec. IV contains the summary of the work.

\section{Model and Theoretical framework}

\subsection{Ring nanojunction and TB Hamiltonian}

Let us begin with the junction setup, shown in Fig.~\ref{fig1}, where a $N$-site nano ring is coupled to source (S) and drain (D) 
electrodes. $I_{\mbox{\small tr}}$ is the transport current which is divided into two arms, upper and lower, of the ring and the arm 
currents are $I_1$ and $I_2$, respectively. The circular current is defined as~\cite{cr1,cr2,cr3}
\begin{equation}
I_c =\frac{I_1 L_1+I_2L_2}{L} 
\label{cireq}
\end{equation} 
where $L=L_1+L_2$ being the circumference of the ring. It is easy to follow that when the two arms are identical (status and lengthwise) 
then $I_1=I_2$ (positive sign is used for clockwise motion) and $ L_1=L_2$. Under that condition $I_c$ will be exactly zero. Thus, in 
order to have a finite $I_c$, we need to break the symmetry between the two arms.

To describe the nanojunction, we employ a TB framework. The setup contains different parts: source, ring, drain and the coupling of the 
ring between two electrodes. For all these parts the TB Hamiltonian can be written in a general form as~\cite{cr1,cr3}
\begin{equation}
H =\sum_\alpha \left[\sum_n \epsilon_{\alpha,n}c^\dagger_{\alpha,n}c_{\alpha,n} + \sum_{n,m} t_{\alpha,n,m} c^\dagger_{\alpha,n}c_{\alpha,m}\right]
\label{eq.1}
\end{equation}
where $\alpha=S,D,R$ respectively for source (S), drain (D), and ring (R). $c^\dagger_{\alpha,n}$, $c_{\alpha,n}$ are the usual 
fermionic operators, $\epsilon_{\alpha,n}$ is the on-site energy and $t_{\alpha,n,m}$ is the nearest-neighbor hopping (NNH) strength. The
index $m$ is used to describe the neighboring sites of the site $n$. Now, we explicitly describe the physical parameters 
$\epsilon_{\alpha,n}$ and $t_{\alpha,n}$ that are associated with different parts of the junction. For the ring, $\epsilon_{\alpha,n}$ 
is denoted by $\epsilon_{R,n}$ and the site energies are chosen in the AAH form governed by the relation~\cite{aa1,aa2,aa3,aa4}
\begin{equation} 
\epsilon_{R,n}= W \cos\left(2\pi b n + \phi_\nu\right).
\end{equation} 
Here $W$ measures the AAH modulation strength or simply called as the correlated disorder strength, $\phi_\nu$ is the phase factor that 
can be tuned externally with a suitable arrangement and $b$ is an irrational number which makes the site energies non-uniform but 
correlated (non-random). For rational $b$ (viz, $b=1/Q$, $Q$ being an integer) and if the total number of lattice sites $N$ of the 
ring is divisible by the factor $Q$ then the ring becomes an ordered one with $Q$ periodicity. For a perfect ring, the phenomenon of 
bias induced circular current is quite known, and thus, in the present work we concentrate on the disordered one. For $W=0$, the ring 
is always perfect irrespective of $b$ and this is a trivial case. The other factor $t_{R,n,m}$ in the ring is associated with electron 
hopping. In our case, we choose two different hopping strengths that repeat one after the other, following the SSH model. These two 
hopping strengths are denoted as $t_1$ and $t_2$. Both $t_1<t_2$ and $t_1>t_2$ conditions are imposed in our calculations for the
dimerized ring, and, for $t_1=t_2$ we have the ring with conventional uniform hopping. The side attached electrodes are taken as 
one-dimensional perfect and reflection-less. The electrodes are parameterized by the on-site energy $\epsilon_0$ and NNH strength 
$t_0$. In our junction setup the source electrode is always coupled to site $1$ of the ring, whereas drain electrode is attached 
to any arbitrary point $q$ (which is variable). The coupling strengths of these electrodes S and D, to the ring are denoted 
by the parameters $\tau_s$ and $\tau_d$ respectively.

\subsection{Theoretical Framework}

To determine circular current inside the ring we have to find bond currents and for that we need to calculate current densities in 
individual bonds. For any particular bond, the current density is defined as~\cite{cr1,cr3,wg1}
\begin{equation}
J_{n,n+1}(E) = \left(\frac{2e}{\hbar}\right) \mbox{Im}\left[t_{R,n}A^*_{R,n}A_{R,n+1}\right]
\label{eq.2}
\end{equation}
where $t_{R,n}= t_1$ or $t_2$ depending on the bond of the ring, $A_{R,n}$'s are the wave amplitudes at different lattices sites, 
$e$ and $\hbar$ ($\hbar = h/2\pi$) are the fundamental constants. Solving $(N+2)$ coupled equations involving wave amplitudes of 
different lattice sites, the coefficients $A_{R,n}$'s are computed. Among which $N$ equations are associated with $N$ lattice sites 
of the ring, and $2$ other equations are related to the sites of the source and drain with which the ring is coupled. This is a 
standard prescription and has been actively used in the literature. The general form of $(N+2)$ equations is expressed as~\cite{wg1}
\begin{equation}
\left(E-\epsilon_{\alpha,n}\right)A_{\alpha,n} = \sum_m t_{\alpha,m}A_m
\label{eq.3}
\end{equation}
where the index $m$ runs over all the nearest-neighbor sites of the site n.

Integrating the bond current density $J_{n,n+1}(E)$ over a suitable energy window, the bond current is calculated. 
It is expressed as~\cite{cr1,cr3,wg1}
\begin{equation}
I_{n,n+1} = \int_{E_F-eV/2}^{E_F+eV/2} J_{n,n+1}(E)\,dE.
\label{eq.4}
\end{equation}
Determining the currents in all the bonds of the upper and lower arms, the net circular current in the ring is evaluated. 
It is written as~\cite{cr1,cr3}
\begin{equation}
I_c = \frac{1}{N}\sum_n I_{n,n+1}.
\label{eq.5}
\end{equation}

\section{Numerical results and discussion}

This section contains all the essential results of our investigation, which include current density and circular currents under different 
input conditions of the junction setup. Before starting our discussion let us mention the parameter values that are kept unchanged 
throughout the work. For the side-attached electrodes we choose $\epsilon_0=0$. The ring-to-electrode coupling strengths $\tau_s$ and 
$\tau_d$ are fixed to $1\,$eV. Unless specified, we choose $N = 10$, $W = 0.5$, $\phi_\nu = 0$, $E_F = 0$, $t_0=2$
and we compute the results for a symmetric junction configuration setting the temperature to zero. The other physical parameters 
that are different compared to the stated above, are mentioned in the appropriate parts. All the energies are measured in units of 
electron-volt (eV). Depending on the condition imposed on $t_1$ and $t_2$, we have two different kinds of ring systems. 
\begin{figure}[ht]
\centering \resizebox*{6.5cm}{9cm}{\includegraphics{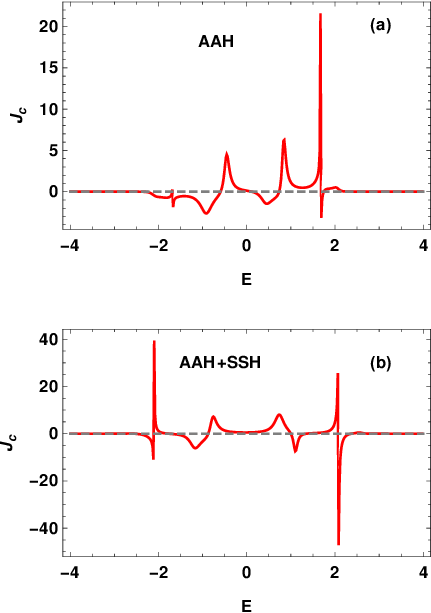}}
\caption{(Color online). Total circular current density $J_c$ as a function of energy $E$ for the (a) AAH ring with $t_1=t_2=1$ and 
(b) AAH SSH ring with $t_1=1.5$ and $t_2=1$. The source and drain are connected to sites $1$ and $6$ of the ring, respectively.}
\label{fig2}
\end{figure}
In one case we 
consider a ring with only correlated disorder which is referred to as AAH ring, and, for the other case both AAH modulation and 
hopping dimerization are present which we may call, for simplification, as AAH SSH ring.

Now we present and analyze the results one by one. Let us begin with Fig.~\ref{fig2} where the variation of total current density 
$J_c$ with incoming electron energy $E$ is shown for the (a) AAH and (b) AAH SSH ring junctions. The total current density is obtained
by adding the current densities associated with all the individual bonds of the ring. For both the ring nanojunctions, peaks and dips 
are observed at some discrete energies and in some energies they are very sharp, while for other energy regions $J_c$ becomes zero. 
These peaks and dips appear at the discrete energy eigenvalues of the ring which is clamped between the electrodes. Unlike the current 
density determined at the drain electrode, which provides only one sign, the current density measured inside the ring can have positive
and negative signs, as electrons can move both in the clockwise and counterclockwise directions. 
\begin{figure}[ht]
\centering \resizebox*{8cm}{4.75cm}{\includegraphics{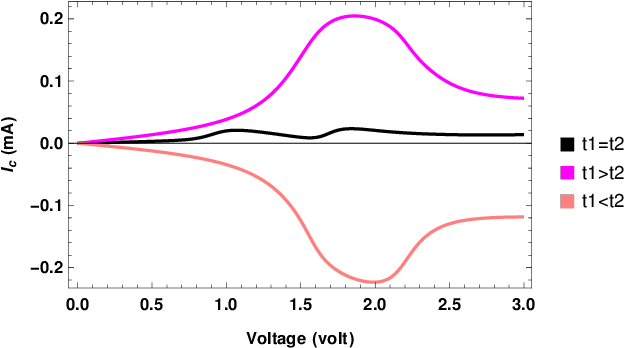}}
\caption{(Color online). Variation of circular current $I_c$ with applied bias voltage $V$ for three different conditions of the 
hopping integrals $t_1$ and $t_2$, where the black, magenta, and pink curves correspond to $t_1=t_2$ ($t_1=t_2=1$), $t_1>t_2$ 
($t_1=1.5$, $t_2=1$), and $t_1<t_2$ ($t_1=1$, $t_2=1.5$), respectively.}
\label{fig3}
\end{figure}
Quite an irregular shape of the density 
envelop across a peak or dip arises due to the disorder in the ring. For a perfect ring ($W = 0$), uniform peaks and dips are obtained 
(not shown here in the figure). Comparing the results given in Figs.~\ref{fig2}(a) and (b) it is found that much higher peaks/dips are 
obtained once we include hopping dimerization in the AAH ring. Naturally, higher circular current is expected for the AAH SSH ring compared 
to the AAH one. It can be visualized from the forthcoming results. This enhancement of peak/dip heights is entirely due to the inclusion 
of specific restriction in hopping integrals.

Figure~\ref{fig3} displays the variation of circular current $I_c$ as a function of bias voltage $V$ under three different conditions of
the hopping strengths $t_1$ and $t_2$, those are represented by three colored curves. For the dimerized case we take both the situations
viz, $t_1<t_2$ and $t_1>t_2$. The nature of $I_c$ is directly connected with $J_c$, as integrating the current density profile over a 
suitable energy window (see Eq.~\ref{eq.4}), $I_c$ is obtained. A finite $I_c$ appears when the peaks and dips of $J_c$ asymmetric over
the selected energy window. If they are symmetric then the net current becomes zero due to their mutual cancellation. For the AAH ring 
with identical hopping ($t_1=t_2$), the contributions from peaks and dips are quite comparable and thus small circular current is 
obtained throughout the chosen bias window (black curve of Fig.~\ref{fig3}). But once the dimerized hopping is imposed, the circular 
current becomes reasonably large compared to the only AAH ring, and it is clearly reflected by comparing the curves  given in 
Fig.~\ref{fig3}. This enhancement of current is solely due to the modification of the current density profile in presence of the 
dimerized hopping. In addition, we find that depending on the condition i.e., whether $t_1>t_2$ or $t_1<t_2$, there is a sign reversal 
of current. The enhancement of $I_c$ by introducing hopping asymmetry ($t_1 \neq t_2$) can be referred to as ``delocalization" phenomenon 
which is of course a new observation in the context of bias driven circular current and might be useful in different other directions.

Following the above fact that $I_c$ gets significantly modified with the hopping factors, 
\begin{figure}[ht]
\centering \resizebox*{8cm}{4.5cm}{\includegraphics{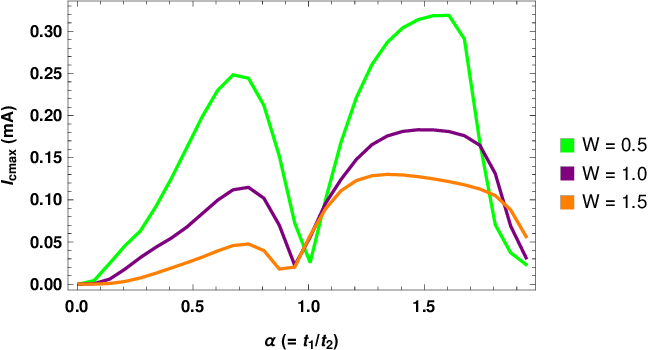}}
\caption{(Color online). Dependence of maximum circular current $I_{cmax}$ on the
ratio of $t_1$ and $t_2$ for three distinct values of disorder strength $W$. Here $t_2$ is fixed at $1.5$ and $t_1$ is varied from
the $0$ to $3$. To have all the energy levels of the ring inside the allowed energy window of the electrodes here we set $t_0=3$. 
The rest other physical parameters and the junction configuration remain unchanged as used in Fig.~\ref{fig2}.}
\label{fig4}
\end{figure} 
\begin{figure}[ht]
\centering \resizebox*{8cm}{5.5cm}{\includegraphics{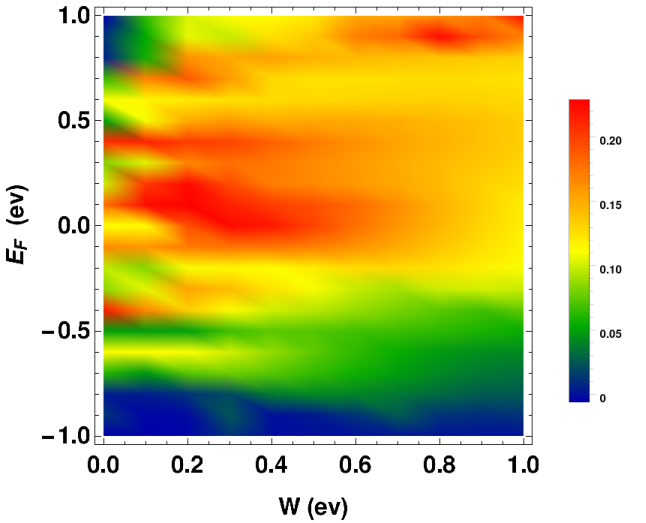}}
\caption{(Color online). Density plot. Simultaneous variation of maximum circular current $I_{cmax}$ as functions of Fermi energy $E_F$ 
and AAH modulation strength $W$ for the AAH SSH ring. Here we take $t_1=1.5$ and $t_2=1$.The color bars denote $I_{cmax}$ in units of mA.}
\label{fig5}
\end{figure}
it is now necessary to check how the current 
is altered if we continuously vary anyone of the two hopping strengths in a broad range keeping the other constant. 
The results are shown in Fig.~\ref{fig4} for three typical values of AAH modulation strength $W$. Here we essentially focus on the current
magnitude, and hence, we plot the absolute maximum of $I_c$ (referred to as $I_{cmax}$) computing $I_c$ in a large bias window ($0$ to 
$3\,$V) and taking the maximum from that. The effect of $\alpha (= t_1/t_2)$ is really appreciable. The circular current can significantly 
be modified by altering the hopping strength and the underlying mechanism solely relies on the modification of energy eigenspectrum of the 
ring and thus the current density profile. In each $W$, the circular current becomes too low for the ring with with only AAH modulation 
(viz, $\alpha = 1$) compared to the cases when $t_1 \neq t_2$, and this effect becomes more prominent for lower values of $W$. It gives 
clear indication that hopping dimerization can lead to a {\em delocalization} in a disordered system. 

All the above results are worked out for a particular value of Fermi energy $E_F$ 
\begin{figure}[ht]
\centering \resizebox*{8cm}{4.625cm}{\includegraphics{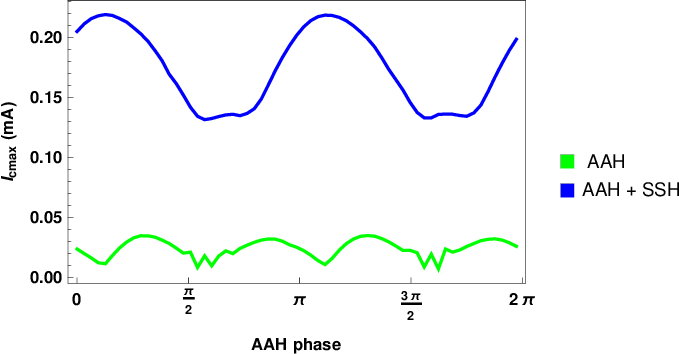}}
\caption{(Color online). Dependence of $I_{cmax}$ on the AAH phase factor $\phi_{\nu}$ both for the AAH ($t_1=t_2=1$) and 
AAH SSH ($t_1=1.5$ and $t_2=1$) rings.}
\label{fig6}
\end{figure}
\begin{figure}[ht]
\centering \resizebox*{8cm}{4.5cm}{\includegraphics{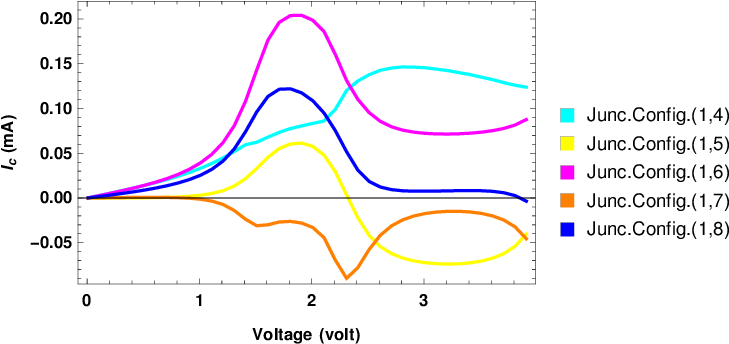}}
\caption{(Color online). Effect of ring-electrode junction configuration on $I_c$. Variation of $I_c$ as a function of bias voltage for
some distinct ring-electrode junction configurations in the AAH SSH ring with $t_1=1.5$ and $t_2=1$.}
\label{fig7}
\end{figure}
and for some typical disorder strengths. Now to inspect
the characteristics of $I_c$ more clearly, in Fig.~\ref{fig5}, we present a density plot where the variation of $I_{cmax}$ is shown by
simultaneously changing $E_F$ and $W$ in a wide range. The results are computed for the symmetric junction configuration. From the 
density plot it is clearly seen that there are multiple Fermi energies for which the current get enhanced with increasing $W$. The 
other important finding is that for a broad range of $E_F$ we can have the possibility to get large $I_c$ which suggests that our results 
are not valid only for a specific choice of $E_F$.

Figure~\ref{fig6} displays another interesting behavior. The dependence of $I_{cmax}$ with the AAH phase factor $\phi_{\nu}$ is shown for 
the two rings, AAH and AAH SSH. In both the cases $I_{cmax}$ varies with $\phi_{\nu}$, and the effect becomes more prominent for the 
AAH SSH ring. The change of $\phi_{\nu}$ alters the site energies of the ring which directly modifies the energy eigenvalues and the 
current density profile, resulting in a variation in $I_c$. The appearance of higher circular current in presence of the hopping 
dimerization is clearly visible.

It is well known that for a ring nanojunction, the ring-electrode junction configuration plays an important role 
\begin{figure}[ht]
\centering \resizebox*{7cm}{4.5cm}{\includegraphics{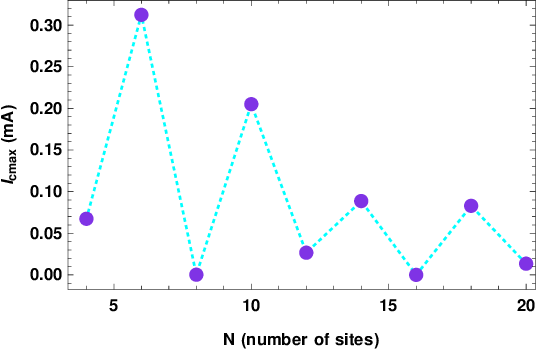}}
\caption{(Color online). Variation of $I_{cmax}$ as a function of ring size $N$ for the lengthwise symmetric junction configuration. 
The ring size is varied from $4$ to $20$ with the interval $\Delta N=2$. The hopping parameters are: $t_1=1.5$ and $t_2=1$.}
\label{fig8}
\end{figure}
on transport phenomena, and thus, its effect on circular current needs to be discussed. In Fig.~\ref{fig7} we present the results 
of $I_c$ as a function of bias voltage for five distinct ring-electrode junction configurations. The current is highly sensitive to 
junction configuration and it is due to the effect of quantum interference of electronic waves passing through different arms of the 
nano ring. The overall current magnitude becomes maximum for the lengthwise symmetric junction (where drain is connected to site number
$6$ of the ring of our chosen $10$-site ring). For other junction configurations the currents are also appreciable.

For the sake of completion of our analysis, finally, we discuss the effect of ring size on circular current. Figure~\ref{fig8} displays 
the variation of $I_{cmax}$ as a function of ring size $N$ by varying it from $4$ to $20$ with the interval $\Delta N = 2$. In each case 
we choose lengthwise symmetric junction. First of all, a finite oscillation is observed with $N$ which is the generic property of the 
low-dimensional quantum systems. The fluctuation is directly connected to the quantum interference of electronic waves. Though for all 
the chosen ring sizes, a finite current is obtained, the overall current envelop shows a decreasing nature with the ring size. This is 
expected as we are discussing a quantum mechanical phenomenon. For large enough ring size the fluctuation will be suppressed significantly 
and the net current will be vanishingly small which we confirm through our detailed numerical analysis. Thus, smaller ring sizes are 
highly recommended to have finite bias-driven circular current.

\section{Summary}

To summarize, the present work deals with the interplay between quasi-periodic site energies and hopping dimerization on bias-driven 
circular current in a ring nanojunction. The site energies are taken in a cosine form following the AAH model and the dimerized 
hopping integrals are included in the form of SSH model. Employing a tight-binding framework the results are computed within a 
wave-guide theory. The hopping dimerization has a strong effect on current enhancement which we establish critically by analyzing current 
under different input conditions. The specific roles of disorder, Fermi energy, ring size and junction configurations on $I_c$ are 
thoroughly investigated. The results are valid for a broad range of physical parameters which proves the robustness of our work. 
The present analysis may provide a new insight on localization behavior in disordered lattices in presence of finite correlation 
among the hopping integrals.
Before an end, we would like to point out that, instead of AAH modulation if we consider random site energies, an enhancement 
of circular might also be expected in the presence of hopping dimerization. However, the response is relatively weak compared 
to the AAH case. 

\end{document}